\documentclass{iopart}
\usepackage{amsfonts}

\begin{document}

\title{On the equation of motion of compact binaries 
in Post-Newtonian approximation
}

\author{Yousuke Itoh}
\address{
Max Planck Institut f\"ur  Gravitationsphysik, Albert Einstein Institut \\ 
Am M\"uhlenberg 1, Golm 14476, Germany
}
\ead{yousuke@aei.mpg.de}

\begin{abstract}
A third post-Newtonian (3 PN) equation of motion 
for two spherical compact stars in a harmonic 
coordinate has been derived based on the 
surface integral approach and the strong field point 
particle limit. 
The strong field point particle limit 
enables us to incorporate a notion of 
a self-gravitating regular star into general relativity.
The resulting 3 PN equation of motion is Lorentz invariant, 
unambiguous, and 
conserves an energy of the binary orbital motion.  
\end{abstract}

\maketitle
\newcommand{\pa}{\partial}

\section{Introduction}

Recently, high order post-Newtonian iteration 
for inspiralling compact binaries has attracted 
a renewed attention in the context of 
the worldwide gravitational waves detection 
efforts \cite{Tagoshi01,Ando01,Takahashi03,LIGOGEOS1IBPaper}. 
Detectability and qualities of measurements 
of gravitational waves emitted by such binaries  depend 
on our theoretical knowledge about waveforms and hence 
partly about dynamics of binaries \cite{CutlerEtAL93}.

The currently available equation of motion and the Hamiltonian 
that govern dynamics of (comparable masses) binaries 
in the post-Newtonian approximation of general relativity are 
the 3.5 PN equations of motion in a harmonic coordinate 
\cite{BF00a,BF01a,PW02}
and the 3.5 PN Arnowitt-Deser-Misner (ADM) Hamiltonian in 
the ADM transverse-traceless (ADMTT) gauge \cite{JS97,JS98,DJS01a}.
For reviews on post-Newtonian iteration, 
we refer readers to, e.g.,  \cite{Blanchet02b}.

In the inspiralling phase of spherical compact stars binaries, 
a point particle description   
is certainly useful and suitable 
to study their dynamics. 
A simple realization of this  
description may be a use of Dirac delta distributions, 
however, it causes ill-defined divergences due to 
the non-linearity of general relativity. The above mentioned 
works (except for \cite{PW02}) have used Dirac delta 
distributions and resorted to regularization procedures 
such as Hadamard's partie finie. Then a problem has occurred.  
At the 3 PN order, 
there exists one arbitrary undetermined parameter 
which reduces the predictability of the theoretical waveforms.
Jaranowski and Sch\"afer \cite{JS98,JS99} who first achieved 
the 3 PN iteration found two undetermined coefficients; 
$\omega_{{\rm k}}$ and $\omega_{{\rm s}}$ in their  
Hamiltonian   
(later they fixed $\omega_{{\rm k}}$ 
by imposing Poincar\'e invariance 
on their Hamiltonian \cite{DJS00}). 
Blanchet and Faye \cite{BF00a,BF01a} 
found one coefficient, called $\lambda$, 
in their equation of motion. 
They then showed \cite{DJS01b,ABF01} 
that the results of these two independent studies 
are consistent with each other when the following relation holds;  
$\omega_{{\rm s}} = -11\lambda/3 - 1987/840$.
It was suggested that  
unsatisfactory features of the regularizations 
they have used associated with their use of Dirac 
delta distributions were   
possible origins of the parameters, 
as summarized in \cite{DJS01a}. 

In fact, Damour, Jaranowski, and Sch\"afer succeeded in 
deriving the complete 3 PN ADM Hamiltonian in the ADMTT 
gauge and fixed the parameter 
as $\omega_{{\rm s}} = 0$ using dimensional 
regularization \cite{DJS01a}. However, 
applicability of mathematical regularizations 
to the current problem is not a trivial issue, 
but an assumption to be verified, or at least supported by 
convincing arguments. 
At the 2.5 PN order, there is a physical argument 
of the ``Dominant Schwarzschild Condition'' \cite{Damour83}  
which supports a use of Dirac delta distributions and a 
certain sort of regularizations.  There is no such argument 
at the 3 PN order. Thus, it seems crucial to achieve  
unambiguous 3 PN iteration without introducing singular 
sources and support the result of \cite{DJS01a}.

The author, Futamase and Asada have recently developed yet 
another method of derivation of a post-Newtonian equation 
of motion in a harmonic coordinate 
appropriate to inspiralling compact binaries 
\cite{IFA00,IFA01} which is based on the strong field 
point particle limit \cite{Futamase87} and the surface 
integral approach \cite{EIH1938}. 
In this paper, due to the lack of space, 
we briefly review our method and present 
few of our 3 PN results.  
Full explanations on our method and results will 
be reported elsewhere \cite{YItoh2004,IF2004}.

\section{Strong field point particle limit}

We shall write the post-Newtonian expansion 
parameter explicitly as $\epsilon$, which represents 
smallness of the orbital velocity, $O(\epsilon)$,  
and weakness of the interstars field,  
$O(\epsilon^2)$. Futamase \cite{Futamase87} 
advocated the strong field point particle limit 
where the radius of 
the star shrinks at the same rate as the mass 
in the post-Newtonian limit ($\epsilon \rightarrow 0$). 
Thus, the typical scale of gravitational field inside 
the star ($\sim$ the mass over the radius of the star) 
remains finite (and can be strong) in $\epsilon$ zero limit. 
This is in contrast with the usual post-Newtonian expansion 
for an extended star, where the matter density is assumed to 
scale as $\epsilon^2$ and we apply the post-Newtonian 
approximation to the field even inside the stars.

\section{Field equations}

Using the post-Newtonian approximation, we iteratively solve the 
Einstein equations under the harmonic gauge condition that is written as  
$h^{\mu\nu}\mbox{}_{,\nu} = 0$ where $h^{\mu\nu} \equiv  \eta^{\mu\nu} 
- \sqrt{-g}g^{\mu\nu}$, $\eta^{\mu\nu} \equiv {\rm diag}(-\epsilon^2,1,1,1)$ 
is the flat metric, and $g$ is the determinant of the metric 
$g_{\mu\nu}$.
Then harmonically relaxed Einstein equations can 
be casted in integral forms as 
$h^{\mu\nu}(\tau,x^i) = 4 \int_{C} d^3y 
\Lambda^{\mu\nu}(\tau-\epsilon|\vec x-\vec y|, y^k; \epsilon) 
/|\vec x-\vec y|$,   
where $\Lambda^{\mu\nu} \equiv \Theta^{\mu\nu} 
+ \chi^{\mu\nu\alpha\beta}\mbox{}_{,\alpha\beta}$,   
$\Theta^{\mu\nu}$ is the matter plus the 
gravitational field stress energy pseudo-tensor 
($\Theta^{\mu\nu} \equiv  
-g(T^{\mu\nu} + t_{LL}^{\mu\nu})$ with $t_{LL}^{\mu\nu}$ 
being the Landau-Lifshitz pseudo tensor) and   
$\chi^{\mu\nu\alpha\beta}\mbox{}_{,\alpha\beta}$ arises since 
we use the wave operator of the flat spacetime instead of 
that of the curved spacetime. 
$\tau$ is the time coordinate.
The light cone $C$ emanating from the field point 
$(\tau,x^i)$ is split into two zones; 
the near zone $N$ which encloses 
the two stars completely 
and the far zone $F$ which is 
the outside of $N$ and where the retardation effects of the field 
are manifest. To treat the retarded integrals over $F$ 
we have adopted the {\it Direct Integration of the 
Relaxed Einstein equations} method developed by Will and collaborators 
\cite{PW02}, however, it is well-known \cite{PW02,BD88} 
that the integrals over $F$ do not contribute to an equation of motion up 
to the 3.5 PN order and hence we devote ourselves to the integrals 
over $N$.    

The integral region $N$ then is split into three regions; 
two spheres, called the body zone $B_A$ ($A=1,2$ labels 
each star), each of which surrounds each star with no intersection 
with the other nor the star,  
and elsewhere $N/B$ where $B \equiv B_1 \cup B_2$.   
We make the radius of $B_A$ shrink proportionally to $\epsilon$ 
to define the multipole moments and to 
derive an equation of motion for compact stars. That is, 
$B_A \equiv \{x^i| |\vec x - \vec z_A(\tau)| \le \epsilon R_A\}$ where 
$\epsilon R_A$ 
is arbitrary but smaller than the orbital separation $r_{12}$ 
and larger than the radius of the star for any $\epsilon$. 
$z_A^i(\tau)$ represents the location of the star $A$  
in the point particle limit. 
The integrals over each $B_A$ are evaluated using the multipole 
expansion of the star. For example, the ($\Theta$ part of the) 
mass multipole moments of the star $A$ are defined as  
$
I_{A\Theta}^{K_l} \equiv 
\epsilon^2 \int_{B_A}
d^3\alpha_A \Theta^{\tau \tau} \alpha_A^{\underline{K_l}}, 
$
where $\alpha_A^{\underline{i}} \equiv 
\epsilon^{-2}(x^i-z_A^i(\tau))$ is a scaled coordinate in which 
the star does not shrink to the point $z_A^i(\tau)$. 
The capital index denotes a set of collective indices, 
$I_l \equiv i_1 \cdot\cdot\cdot i_l$  and
$\alpha_A^{\underline{I_l}} \equiv
\alpha_A^{\underline i_1}\alpha_A^{\underline i_2}
\cdot\cdot\cdot\alpha_A^{\underline i_l}$. 

The remaining Poisson-type integrals over $N/B$, 
$
\int_{N/B} d^3y 
|\vec{x}-\vec{y}|^{n-1}f(\tau,\vec{y}) 
$ 
($n \ge 0 :$ integer, $f(\tau,\vec x)$ being a certain combination of 
$\Lambda^{\mu\nu}(\tau,\vec x)$), are evaluated with the help of 
particular solutions of Poisson equations. For $n=0$ case, 
using $\Delta g(\vec x) = f(\vec x)$ (for brevity, we do not 
write $\tau$ dependence in $f$ nor $g$ here), we have  
$$
\int_{N/B} \frac{d^3y f(\vec{y})}{|\vec{x}-\vec{y}|}
= 
- 4 \pi g(\vec{x}) + \oint_{\pa(N/B)}dS_k
\left[\frac{1}{|\vec{x}-\vec{y}|}
 \frac{\pa g(\vec{y})}{\pa y^k} - g(\vec{y})\frac{\pa}{\pa y^k}
 \left(\frac{1}{|\vec{x} - \vec{y}|}\right) \right].
$$
Once explicit expressions of the necessary particular solutions 
are in hand, it is straightforward to evaluate the Poisson-type 
integrals.

\section{Mass-energy relation and momentum-velocity relation}

In our formalism, we do not assume a specific expression of 
the matter's stress energy tensor. Thus, we need to 
express the operationally defined multipole moments, 
namely, the four momentum 
$
P_{A\Theta}^{\mu} \equiv 
\epsilon^2 \int_{B_A} d^3\alpha_A \Theta^{\mu \tau} 
$ of the 
star in terms of the mass $m_A$, the velocity $\vec v_A$, 
and $\vec r_{12}$.  As for 
$P_{A\Theta}^{\tau}$, we use its evolution equation  
which is derived by the local energy momentum 
conservation $\Theta^{\mu\nu}\mbox{}_{,\nu} =0$ as   
$
dP_{A\Theta}^{\tau}/d\tau = -\epsilon^{-4}
 \oint_{\pa B_A} dS_k (-g)t_{LL}^{k\tau}
+\epsilon^{-4}
v_A^k \oint_{\pa B_A} dS_k (-g)t_{LL}^{\tau\tau}. 
$ 
At the lowest order, we obtain $d P_{A\Theta}^{\tau}/d\tau = 
O(\epsilon^2)$.  The evolution equation 
of $P_{A\Theta}^{\tau}$ is then solved functionally into a 
form $P_{A\Theta}^{\tau} = m_A[1+O(\epsilon^2)]$ where 
the mass of the star $A$, $m_A$, is defined as  
the integration constant by 
$m_A \equiv \lim_{\epsilon \rightarrow 0}P_{A\Theta}^{\tau}$. 
We note that although $m_A$ is defined by $\epsilon \rightarrow 0$ 
limit of $P_{A\Theta}^{\tau}$, the 
gravitational field energy inside the star 
is taken into account in 
the definition of $m_A$ since 
$P_{A\Theta}^{\tau}$ is defined as a volume integral 
whose integrand includes the field 
energy $(-g)t_{LL}^{\tau\tau}$ which does not vanish  
in the $\epsilon$ zero limit because of the scaling 
of the strong field point particle limit.   
(the mass over the radius of the star  
is finite inside the star independently of $\epsilon$). 
Following the Newtonian dynamics, 
the momentum-velocity relation on the other hand is 
derived by the time derivative of the 
dipole moment $D_{A\Theta}^i \equiv 
\epsilon^2 \int_{B_A} d^3\alpha_A \Theta^{\tau \tau} 
\alpha_A^{\underline i}$ as
$P^{i}_{A\Theta} = P^{\tau}_{A\Theta} v^i_A + Q_{A\Theta}^i +
\epsilon^2 d D_{A\Theta}^i/d\tau$ where 
$Q_{A\Theta}^i \equiv \epsilon^{-4}
\oint_{B_A}dS_k((-g)t_{LL}^{\tau k}-v_A^k(-g)t_{LL}^{\tau\tau})y_A^i
 = -\epsilon^6 d (m_A^3 \mbox{}_0a_{A}^i/6)/d\tau
$ 
($\mbox{}_0a_{A}^i$ is the Newtonian acceleration) 
arises due to the non-compactness of $(-g)t_{LL}^{\mu\nu}$.
$Q_{A\Theta}^i$ 
appears starting from the 3 PN order and 
it affects the 3 PN equation of motion.
Finally, we note that as in the Newtonian case, we can choose 
the value of the dipole moment $D_{A\Theta}^i$ freely to define 
the representative point of the star $A$, $z_A^i(\tau)$. 
For instance, $z_A^i(\tau)$ corresponding to $D_{A\Theta}^i = 0$ 
may be called the center of mass of the star $A$. 
However, we found that it is convenient to adopt 
$D_{A\Theta}^i = \epsilon^4 (1/6 - 22\ln(r_{12}/\epsilon/R_A))
m_A^3 \mbox{}_0a_{A}^i$ at the 3 PN order \cite{YItoh2004,IF2004}. 
Choosing a certain value for $D_{A\Theta}^i$ corresponds 
to choosing the corresponding coordinate under the harmonic gauge 
condition.  Our particular choice gauges $\ln \epsilon R_A$ 
dependence away from the 3 PN acceleration. 
A different choice leaves $\ln \epsilon R_A$ dependences in
an acceleration, however, these are mere gauge 
terms and do not appear when we are concerned with coordinate 
invariant quantities. In fact, a different choice 
$D_{A\Theta}^i = \epsilon^4 m_A^3 \mbox{}_0a_{A}^i/6$ gives 
the same invariant binary orbital energy of circular orbital motion 
in the center of mass frame as $E(x)$ given below \cite{IF2004}.

\section{Equation of motion}

There may be three methods being studied in  
derivation of a high order post-Newtonian equation of motion.
The first one is the volume integral approach, which may take 
the following form in the Newtonian dynamics:  
$m_A d v_A^i/d\tau = \int d^3y 
\rho_A(\vec y) \phi(\vec y)^{,i}$. 
($\rho_A$ denotes the density of the star $A$. 
$\phi$ is the Newtonian potential.)
In this method, one can track how the stars' 
internal structures dependent terms cancel out (or remain) 
and disappear (or appear) in the final result; 
if the resulting equation 
of motion depends on, say, the density profile of 
(almost stationary) spherically symmetric compact stars, 
it suggests violation of the equivalence principle to 
some extent.  This method is being used by 
Pati and Will \cite{PW02}.

The second method assumes that a regularized 
action describes dynamics of binaries. 
Blanchet and Faye \cite{BF00a,BF01a} have used this 
method with the generalized Hadamard's partie finie 
regularization but obtained one arbitrary parameter $\lambda$ 
in their 3 PN equation of motion. 
It is shown \cite{BF00a,BF01a} that a regularized geodesics can be 
derived from a certain action. This method is attractive, 
besides 
its mathematical beauties,  
because if the resulting equation coincides with 
the equations derived by the first (and the third below) method,  
then it suggests a notion of 
``geodesics in a dynamical spacetime'' 
(with a certain class of regularizations supplemented).

The third method, which we have adopted, is the surface 
integral method where we compute the gravitational 
momentum flux $t^{ij}$ going through a sphere surrounding 
the star concerned.  In the Newtonian case, we may have 
$m_A d v_A^i/d\tau = - \int dS_j t^{ij}$;     
$
t^{ij} = 1/(4\pi)
(\phi^{,i}
\phi^{,j}
-\delta^{ij}
\phi^{,k}\phi_{,k}/2).
$
In this method, possible information of 
the stars' internal structures are coded in, 
say, the multipole moments of the stars and hence 
it affects the equation of motion through $t^{ij}$.  
Based on the local energy momentum conservation, 
we obtain in our formalism 
a simple generalization of the surface 
integral approach to general relativity as 
\begin{eqnarray}  
m_A \frac{d v_A^i}{d\tau}
&=& - 
\oint_{\pa B_A}dS_k (-g)t_{LL}^{ki} 
+ v_A^k 
\oint_{\pa B_A}dS_k (-g)t_{LL}^{\tau i}
\nonumber \\
\mbox{} &+& 
(m_A - P_{A\Theta}^{\tau})\frac{d v_A^i}{d\tau}
+ \frac{dP_{A\Theta}^{\tau}}{d\tau} v_A^i
- \frac{d Q_{A\Theta}^i}{d\tau} 
- \frac{d^2 D_{A\Theta}^i}{d\tau^2}.     
\label{generaleom3PN}
\end{eqnarray}

\section{Gravitational field and equations of motion up to the 3 PN order}

Up to the 2.5 PN order, it is possible to derive the 
gravitational field explicitly by the method described above. 
Once we have the field in closed form, it is straightforward to evaluate the 
surface integrals in Eq. (\ref{generaleom3PN}). 
However, it does not seem possible to find explicitly all the  
particular solutions for the Poisson-type integrals over $N/B$ 
necessary to evaluate the 3 PN gravitational field, even in the
neighborhood of the 
star. We thus partly abandon derivation of the field and 
took an alternative method \cite{YItoh2004}. 
We change the order of integration; 
for the integrands for which we could not find the particular 
solutions of Poisson equations, 
we first evaluate the surface integrals in Eq. 
(\ref{generaleom3PN}) before evaluating the Poisson-type integrals 
and then evaluate the remaining volume integrals. With this method, 
we obtain a 3 PN equation of motion. 
Because of the enormous length of the final result, 
we here present only the 3 PN relative acceleration 
in the case of the circular orbit 
and in the center of mass frame, which is an appropriate equation 
to inspiralling binaries;  
$d V^i/d\tau = - \Omega^2 r_{12}^i + \epsilon^5 
\mbox{}_{{\rm 2.5 PN}}A^i$  
where $V^i = v_1^i - v_2^i$ is the relative velocity 
and  
$\mbox{}_{{\rm 2.5 PN}}A^i$ is the relative 
acceleration at 
the 2.5 PN order (the radiation reaction term). 
The 3 PN orbital angular frequency $\Omega$ is,  
\begin{eqnarray}
m^2 \Omega^2 &=& 
\gamma^3\left[
1 + \epsilon^2\gamma(-3+\nu)
+\epsilon^4\gamma^2 \left(6+\frac{41}{4}\nu + \nu^2\right)
+\epsilon^6\gamma^3 \left(-10
\right.  \right. 
\nonumber \\
\mbox{} &+&
\left. \left. 
 \nu\left\{
-\frac{2375}{24}+\frac{41\pi^2}{64}
\right\}
+\frac{19}{2}\nu^2 + \nu^3 
\right)
\right]
+ O(\epsilon^7),
\label{eq:3PNorbitalangularfrequency} 
\end{eqnarray}
where 
$m = m_1 + m_2$, $\nu = m_1m_2/m^2$, $\gamma = m/r_{12}$. 
We here note that it is not allowed to 
fix the $\lambda$ parameter by 
comparing  
Eq. (\ref{eq:3PNorbitalangularfrequency}) 
with the corresponding result of 
Blanchet and Faye \cite{BF00a}, since  
the harmonic gauge condition both groups have used 
does not fix a coordinate completely \cite{IF2004,YItoh2004}.

Although we can not show it here due to the lack of 
space, our equation of motion respects the Lorentz invariance. 
And more importantly, our 3 PN equation of motion 
is unambiguous and admits a 
conserved energy. In the case of the circular orbit  
(neglecting the radiation reaction force) and in the center of 
mass frame, the invariant orbital energy $E$ becomes 
$$
E(x) 
=  
- \frac{m \nu x}{2}
\left[
1 + \epsilon^2\left(-\frac{3}{4}-\frac{1}{12}\nu\right)x 
+ 
\epsilon^4
\left(-\frac{27}{8}+\frac{19}{8}\nu - \frac{1}{24}\nu^2\right)
x^2 
\right.
$$
$$
\left. 
+
\epsilon^6
\left(
- \frac{675}{64} + 
\left\{
\frac{34445}{576}-\frac{205\pi^2}{96}
\right\}\nu
-\frac{155}{96}\nu^2
-\frac{35}{5184}\nu^3
\right)
x^3
\right] 
+ O(\epsilon^7),
$$
where  $x = (m\Omega)^{2/3}$. 
We have thus derived a 3 PN equation of motion 
free from an arbitrary parameter. 
Since we introduced $R_A$ in the definition of $B_A$, 
it may seem that $R_A$ appears in the field and also 
in the equation of motion. 
$B_A$ is introduced to separate the region 
where the presence of the star may make the field strong 
from well outside of the stars where the post-Newtonian 
approximation to the field is applicable. The contribution 
to the field from $B_A$ is evaluated with the multipole 
expansion. The equation of motion is derived via the 
surface integrals over $\partial B_A$. 
Now, as for $R_A$ dependent terms in the field, since we have used the 
non-singular matter sources, the field is smooth near 
$\partial B_A$ and possible $R_A$ dependences in the 
body zone contributions to the field cancel out the 
corresponding $R_A$ dependences in the $N/B$ contributions. 
As for an equation of motion, we have proved  in \cite{IFA01} 
that an equation of motion derived from Eq. (\ref{generaleom3PN}) 
does not depend on $R_A$.
See \cite{IFA00,IFA01} for more details. 

Comparing our $E(x)$ with that in \cite{BF00a}, 
we determine 
the coefficient undetermined in the Blanchet and 
Faye 3 PN equation of motion as 
$\lambda = -1987/3080$. This value of $\lambda$ 
is consistent with the result of Damour, Jaranowski, 
and Sch\"afer \cite{DJS01a}. Thus, our result 
(indirectly) 
validates 
their use of dimensional regularization in the 
ADM Hamiltonian approach in the ADMTT gauge.

In the paper \cite{YItoh2004,IF2004}, we devote ourselves 
to two spherical compact binaries, however, our formalism 
can be extended to include higher order multipole moments 
of the stars \cite{IFA00}. 
We have used the strong field point particle 
limit and the surface integral approach to treat strongly 
self-gravitating stars.
Therefore our 3 PN equation of motion is applicable to 
regular compact stars with strong self-gravity.
As is clear from Eq. (\ref{eq:3PNorbitalangularfrequency}), 
our equation of motion depends only on $m_A$, and this 
fact supports the strong equivalence principle up to 
the 3 PN order.

Finally, as mentioned above, we could not derive the 
complete 3 PN gravitational field in closed forms. This fact 
makes it difficult to proceed to the 4 PN iteration. 
The enormous numbers of terms in the intermediate 
computations, of order of $10^5$, appear even at the 
3 PN order. Brute force post-Newtonian iteration 
at the 4 PN order may need efficient computations 
techniques. It may be possible to adopt the strong field 
point particle limit in the currently available waveform 
computation formalism and fix the undetermined parameters 
in the 3.5 PN accurate waveforms.

\ack
The author would like to acknowledge T. Futamase and 
H. Asada for fruitful   
discussions and comments.

\end{document}